\documentclass[twocolumn,amsmath,prl,superscriptaddress,nobibnotes,showpacs]{revtex4}
\usepackage{graphicx}

\begin{document}
\title{Identifying entanglement using quantum ``ghost" interference and imaging}
\author{Milena D'Angelo}
\email{dmilena1@umbc.edu} \affiliation{Department of Physics,
University of Maryland, Baltimore County, Baltimore, Maryland,
21250}

\author{Yoon-Ho Kim}
\email{yokim@umbc.edu} \affiliation{Center for Engineering Science
Advanced Research, Computer Science \& Mathematics Division, Oak
Ridge National Laboratory, Oak Ridge, TN 37831}

\author{Sergei P. Kulik}\altaffiliation[Permanent address: ]{Department of Physics, Moscow
State University, Moscow, Russia.} \affiliation{Department of
Physics, University of Maryland, Baltimore County, Baltimore,
Maryland, 21250}

\author{Yanhua Shih}
\affiliation{Department of Physics, University of Maryland, Baltimore
County, Baltimore, Maryland, 21250}

\date{January 21, 2004}

\begin{abstract}

We report a quantum interference and imaging experiment which quantitatively demonstrates that
Einstein-Podolsky-Rosen (EPR) type entangled two-photon states exhibit both momentum-momentum and
position-position correlations, stronger than any classical correlation. The measurements show indeed that the
uncertainties in the sum of momenta and in the difference of positions of the entangled two-photon satisfy both
EPR inequalities $\Delta(k_{1}+k_{2})<min(\Delta k_{1},\Delta k_{2})$ and $\Delta(x_{1}-x_{2})<min(\Delta
x_{1},\Delta x_{2})$. These two inequalities, together, represent a non-classicality condition. Our measurements
provide a direct way to distinguish between quantum entanglement and classical correlation in continuous
variables for two-photons/two photons systems.
\end{abstract}


\maketitle

The concept of multi-particle quantum entanglement, one of the
most surprising consequences of quantum mechanics, was introduced
in the very early days of quantum theory \cite{schro, EPR}. Since
the development of spontaneous parametric down-conversion (SPDC)
as an efficient source of two-photon entangled states in late
1980's \cite{Klyshkobook}, many experiments have been realized to
exhibit and, afterwards, to exploit the very surprising quantum
effects of entangled states for secure communication, information
processing, and metrology applications \cite{applications}.

Some of the most intriguing effects of two-photon entanglement in SPDC are quantum `ghost' interference and
imaging \cite{Ghost_d,Ghost_i}. These effects are of great importance in potential applications like quantum
metrology and lithography \cite{metro, Dowling,Milena}. Recently, it has been claimed that the two-photon
`ghost' image can be achieved using a pair of classically {\bf k}-vector correlated optical pulses \cite{Bob}.
Ref.~\cite{Bob}, therefore, raises interesting questions about fundamental issues of quantum theory, namely: (i)
to what extent can quantum entanglement in continuous variables be simulated with classically correlated
systems? and (ii) can we experimentally make a distinction between them?

In this Letter, we report an experiment which sheds light on these two tightly related questions. Our idea is to
exploit quantum interference-imaging effects to verify experimentally the EPR-type inequalities, which allow
distinguishing quantum entanglement from classical correlation in continuous variables, for two-photon systems.
By analyzing the results of a two-photon interference and imaging experiment, we show quantitatively that
\textit{entangled} two-photon pairs exhibit both momentum-momentum and position-position EPR-type correlations,
which are stronger than any classical correlation. In contrast, pairs of particles having a perfect
\textit{classical correlation} in momentum (or position), cannot exhibit any correlation in position (or
momentum), due to the uncertainty principle. Our experiment, therefore, shows that entanglement in momentum and
position variables for two-photon systems can be verified experimentally and suggests that the degree of
entanglement can be quantified through the EPR-type non-classicality conditions. This result is of particular
interest, since it represent a ``Bell's inequality" for continuous variables, for two-photon systems.

%%%%%%%%%%%%%%%%%%%%%%%%%%%%%%%%%%%%%%%%%%%%%%%%%%%%%%%%
Consider a pair of EPR-correlated particles. As pointed out by EPR, the most peculiar characteristic of
EPR-entanglement is its independency on the selected basis: entanglement in momentum automatically implies
entanglement in position. In EPR notation \cite{EPR}, the quantum state of entangled two-particle pairs can
indeed be written as
\begin{equation}\label{eprst0}
\Psi(x_1,x_2)=\int{u_p(x_1) \psi_p(x_2)dp}=\int{v_x(x_1)
\phi_x(x_2)dx},\nonumber
\end{equation}
where $x_1$ ($x_2$) is the variable used to describe particle 1
(particle 2), $u_p(x_1)$ ($\psi_p(x_2)$) is the momentum
eigenfunction for particle 1 (particle 2), and $v_x(x_1)$
($\phi_x(x_2)$) is the corresponding position eigenfunction
obtained by Fourier transform of $u_p(x_1)$ ($\psi_p(x_2)$).

As suggested by EPR, an important consequence of entanglement
appears explicitly by considering the case in which $u_p(x_1)$
($\psi_p(x_2)$) is a plane wave. In this case the EPR entangled
state assumes an interesting form:
\begin{eqnarray}\label{eprst}
\Psi(x_1,x_2)=\int{\delta(p_1+p_2) e^{i p_1 x_1/\hbar} e^{i p_2 x_2/\hbar} dp_1 dp_2}=\\\nonumber =h
\delta(x_1-x_2).
\end{eqnarray}
A perfectly EPR-correlated particle pair should therefore be characterized by both the values
$\Delta(p_1+p_2)=0$ and $\Delta (x_1- x_2)=0$. However, even when these uncertainties are different from zero
(non-perfect entanglement), they still satisfy the inequalities:
%%%%%%%%%%%%%%%%%%%%
\begin{eqnarray}\label{ineq}
\Delta (p_1+ p_2) < min(\Delta p_1,\Delta p_2), \\ \nonumber \Delta (x_1- x_2) < min(\Delta x_1,\Delta x_2).
\end{eqnarray}
%%%%%%%%%%%%%%%%%%%%%
The whole entangled system is indeed described in such a way that both the sum of the momenta and difference in
the positions are known with a high degree of accuracy even if the momentum and position of each particle are
completely undefined \cite{EPR}. This result comes directly from the \textit{coherent superposition} of
\textit{two-particle amplitudes}, which cannot be achieved by any classically correlated pairs of particles, as
discussed in more details later.

We propose Eq.~(\ref{ineq}) as a non-classicality condition for continuous variables, for
two-photon/two-particle systems. We report an experimental verification of Eq.~(\ref{ineq}) which exploits
quantum `ghost' interference and image effects of entangled two-photon pairs. We measure $\Delta({\bf k} _1+{\bf
k} _2)$ from a quantum interference experiment and $\Delta({\bf r} _1-{\bf r} _2)$ from a quantum imaging
experiment, both realized using the same SPDC source. To the best of our knowledge a direct quantitative
verification of Eq.~(\ref{ineq}) for two-photon system, i.e. for real momentum and position variables, has not
been reported in literature \cite{squeezing}.

%%%%%%%%%%%%%%%%%%%%%%
\begin{figure}[t]
\includegraphics [width=3.2in]{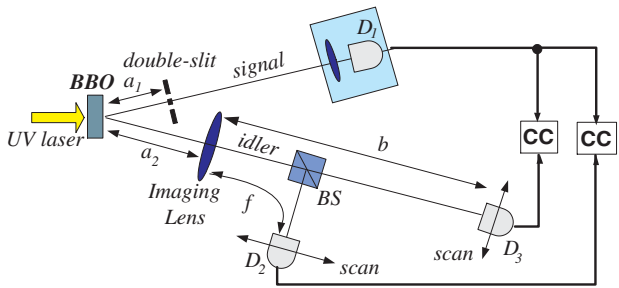}
\caption[] {\label{setup_yoon} Schematic of the experimental setup for observing the two-photon `ghost'
interference and `ghost' image. For simplicity, the prism to remove the pump and the polarizing Thompson prism
are not shown. The double-slit has width $a=0.165$ mm and slit distance $d = 0.4$ mm. The imaging lens has focal
length $f=510$ mm. Relevant distances in this experiment are $a_1=32.5$ cm, $a_2=46.5$ cm, and $b=142$ cm.}
\end{figure}
%%%%%%%%%%%%%%%%%%%%%

Let us first examine whether SPDC two-photon pairs would really
exhibit EPR-type entanglement. Under the assumption that the pump
beam is a plane wave and the transverse dimensions of the pump
beam and the down-conversion crystal are much bigger than the
wavelengths of the photons, the quantum state of the SPDC
two-photon pairs can be written as \cite{Rubin,Rubin_tr}:
\begin{equation}
\label{spdc}
|\Psi\rangle=\sum_{s,i}\delta(\omega_s+\omega_i-\omega_p) \,
\delta ( {\bf k}_{s}+{\bf k}_{i}-{\bf k}_p )\, a_{s}^{\dagger }\,
a_{i}^{\dagger }\,|0\rangle,
\end{equation}
where $\omega _j$ and ${\bf k}_j$ (with $j = s, i, p$) are the frequency and wavevector of the signal (s), idler
(i), and pump (p), respectively, and $a_{s}^{\dagger }$ ($a_{i}^{\dagger }$) is the creation operator for the
signal (idler) photon. Since in this Letter we are only interested in the transverse correlation of the
entangled two-photon pairs \cite{Rubin_tr}, the quantum state used in our experiment is indeed very close to the
one of the original EPR-type entangled pairs of particles. Verification of non-classicality using
Eq.~(\ref{ineq}) should then be possible through adequate experiments realized with this source.

A schematic of the experimental setup can be seen in Fig.~\ref{setup_yoon}. The $351.1$ nm line of an argon ion
laser is used to pump a BBO crystal cut for type-II collinear degenerate parametric down conversion. Pairs of
orthogonally polarized signal and idler photons at central wavelength $\lambda_i=\lambda_s=702.2$ nm, which are
entangled in momentum, emerge from the crystal almost collinearly with the pump laser. After the crystal, the
pump laser beam is separated from the SPDC beam by a quartz dispersion prism. A polarization beam splitting
Thompson prism separates the co-propagating signal and idler into two separate spatial modes. The signal photon
propagates through a double-slit toward a detector package ($D_1$) consisting of a collection lens ($500$ mm
focal length) and a single photon detector placed in its focus. The idler photon propagates freely before being
collected by the imaging lens ($f = 510$ mm). A 50-50 beam splitter (BS) is inserted after the lens. The
reflected and transmitted photons are then detected by single photon detectors $ D_2 $ and $ D_3 $,
respectively. Each of them is mounted on an encoder driver to scan its own transverse plane. A spectral filter
centered at $702.2$ nm with $3$ nm bandwidth precedes each detector. The output pulses of the detectors are sent
to a coincidence circuit ($CC$). Coincidences are measured between $D_1$ and $ D_2 $ and between $ D_1  $ and $
D_3  $.

%%%%%%%%%%%%%%%%%%%%%%
\begin{figure} [t]
\includegraphics[width=2.5in]{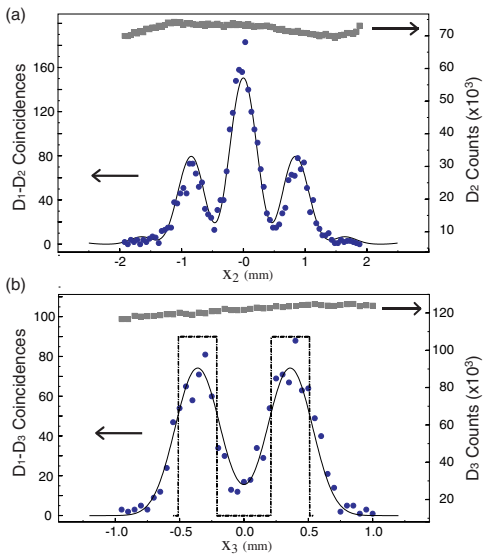}
\caption[] {\label{result} Experimental data. (a) `Ghost' interference-diffraction pattern. (b) `Ghost' image
pattern. $\Delta({\bf k} _1+{\bf k} _2)$ and $\Delta({\bf r} _1-{\bf r} _2)$ are evaluated from each of the
fitting curves (solid lines). The squares are single counts, the dots are coincidence counts.}
\end{figure}
%%%%%%%%%%%%%%%%%%%%%%

This setup therefore allows to measure both `ghost' interference-diffraction and `ghost' image patterns of the
double-slit \cite{Ghost_d,Ghost_i}. Indeed, the \textit{coherent superposition} of biphoton amplitudes allows
exploiting the momentum-momentum correlation to obtain an image (position-position correlation) by simply
changing the observation plane ($D_3$, instead of $D_2$) \cite{Rubin_tr,Rubinnew,Gatti}. The results are shown
in Fig.~\ref{result}. The single counts on both $D_2$ and $ D_3 $, which are scanned in the transverse
direction, show no features at all. The single counting rate of $D_1$, when the detector scans the focal plane
of the collection lens, did not show any interference fringes as well: only a wide bell-shaped pattern was
observed. This result is due to the fact that biphotons are generated with all possible momenta ${\bf k} _i$ and
${\bf k} _s$ such that ${\bf k} _s+{\bf k} _i={\bf k}_p$ is satisfied. In our experiment, the divergence of the
SPDC beam $\Delta(\theta)$, which takes into account the filters bandwidth, the dispersion in the crystal, and
the phase matching condition, is such that: $ \Delta(\theta) \approx 2.6 \textrm{ mrad} \gg \lambda / d, $ where
$d = 0.4$ mm is the distance between the slits and $\lambda=702.2$ nm is the central wavelength of the SPDC
photons. Under this condition, the first order interference-diffraction pattern on $ D_1 $ is simply washed out.

It is, however, possible to observe a `ghost' interference-diffraction pattern when counting coincidences
between $D_1$ and $D_2$ ($D_1$ is fixed and $D_2$ is scanned in the focal plane of the imaging lens) and  to
observe a `ghost' image pattern in coincidences between $D_1$ and $D_3$ ($D_1$ is, again, fixed and $D_3$ is
scanned in the image plane).

For `ghost' interference-diffraction measurement, the detector package $D_1$ plays the role of a point-like
detector. As shown in Ref. \cite{Ghost_d}, we expect the coincidence counting rate to be:
\begin{equation}
\label{Rc} R_c(x_2) \propto {\rm sinc}^{2}[x_2 \pi a/(\lambda f)] {\rm cos}^{2}[x_2 \pi d/(\lambda f)],
\end{equation}
where $x_2$ is the transverse position of detector $ D_2  $ in the focal plane of $L_2 $. Figure~\ref{result}(a)
shows the `ghost' interference measurement. The continuous line in Fig.~\ref{result}(a) is a fitting of the
experimental data, which takes into account both the finite size of the detectors, the divergence of the pump
and the less-than-perfect correlation between signal and idler. It is well known that the visibility of the
interference pattern ($\propto 1 + V {\rm cos}[2 x_2 \pi d/(\lambda f_2)]$) is related to the degree of
transverse coherence of the source. We exploit this effect to evaluate the less-than-perfect transverse
correlation between signal and idler photons:
$$
\Delta(k_{x_s}+k_{x_i})= 2.5 \pm 0.6 \textrm{ mm}^{-1}.
$$
From the divergence of the SPDC beam $\Delta(\theta)$ mentioned above we evaluate $\Delta k_{x_j} \approx 23
\textrm{ mm}^{-1}$, for $j=i,s$. So the ghost interference/diffraction experiment demonstrates that:
$$
\Delta(k_{x_s}+k_{x_i}) \ll min(\Delta k_{x_i},\Delta k_{x_s}),
$$
for our SPDC source.

In a similar way, we obtain $\Delta({\bf r} _s-{\bf r} _i)$ by studying the `ghost' image obtained by measuring
coincidences between $ D_1  $ and $ D_3  $, see Fig.~\ref{result}(b). To observe a `ghost' image, the two-photon
Gaussian thin lens equation, $1/s_i + 1/s_o = 1/f$, where $s_i=b$ and $s_o=a_1+a_2$, should be satisfied
\cite{Ghost_i}. In the ideal situation the detector package $D_1$ is a perfect a bucket detector, which detects
any signal photon that have not been stopped by the double slit. The role of the double slit is then to measure
the localization of the signal with an uncertainty $\Delta(x_s)$, equal to the slit width $a$. In the ideal
case, counting coincidences between $D_1$ and $D_3$, we would obtain two rectangles of width $a' = m a$, and
center-to-center distance $d' = m d$, where $m = s_i/s_o$ is the magnification. In our case $m = 1.8$, $a =
0.165$ mm, $d = 0.4$ mm and the corresponding ideal result ($a' = 0.297$ mm, $d' = 0.72$ mm) is plotted as
dashed line in Fig.~\ref{result}(b). To take into account a more realistic situation we fit the data with the
convolution of the double slit with a Gaussian function that takes into account the finite size of $D_2$. The
comparison of the resulting fitting curve with the theoretical result, dashed line in Fig.~\ref{result}(b),
allows to evaluate $\Delta(x_s-x_i)$, as the difference between the FWHM of the two curves:
$$
\Delta(x_s-x_i) = 0.11 \pm 0.02 \textrm{ mm} \ll \Delta x_s.
$$
Note that the center-to-center distance between the bell-shaped fitting curve and the two rectangles is exactly
the same. The imperfect correlation in position is evidently smaller than the distance between the two slits.

The non-classicality conditions introduced in Eq.~(\ref{ineq}) are then satisfied for the momentum and position
variables of entangled two-photon of SPDC. As we mentioned earlier, this result is a direct consequence, if not
the definition, of quantum correlation: particles that are entangled in momentum are automatically entangled in
position. Only entangled particles can satisfy such inequalities.

An interesting way of understanding this result is the following. The Fourier transform of an entangled state
such as the one given in Eq.~(\ref{eprst}) and (\ref{spdc}), can be factorized by introducing the variables
${\bf k}_1+{\bf k}_2$ and ${\bf k}_1-{\bf k}_2$. The corresponding Fourier transformed variables are ${\bf
r}_1+{\bf r}_2$ and ${\bf r}_1-{\bf r}_2$, respectively. Since ${\bf k}_1+{\bf k}_2$ and ${\bf r}_1-{\bf r}_2$
are not conjugate variables, Eq.s~(\ref{ineq}) can definitely be true simultaneously. For this same reason, also
the product of the two uncertainties can be smaller than one. Indeed we obtain from the uncertainties evaluated
above:
\begin{equation}
\Delta(k_{x_s}+k_{x_i}) \Delta(x_s-x_i) = 0.3 \pm 0.1 < 1,
\end{equation}
Note however that this last inequality is a necessary but not sufficient condition for entanglement \cite{EPR
ineq}.

%%%%%%%%%%%%%%%%%%%%%%%%%%%%%%
Let us now consider the case of two particles or beams classically
correlated in momentum. An example of such a source is given by a
pair of bounded identical guns which emit (quantum) particles
while rotating simultaneously, in such a way that the momenta of
the two particles are always equal in modulus but with opposite
direction. Each pair of \textit{independent but correlated}
particles, fired at a certain angle at a given time, may be
described by:
$$
\left| \Psi_j \right\rangle_{12} = a_{1}^{\dagger }({\bf k}_{j})\
a_{2}^{\dagger }(-{\bf k}_{j})|0\rangle.
$$
If each pair of particles has (non-negative) probability $P({\bf
k} _j)$ of being emitted by the source, the resulting
\textit{incoherent statistical mixture} is described by the
following density matrix:
\begin{equation}
\label{class_2} \rho_{12} = \sum_{{\bf k} _j} P({\bf k} _j) \left| \Psi_j \right\rangle_{12\:12} \left\langle
\Psi_j \right| = \sum_{{\bf k} _j} P({\bf k} _j) \rho_{1}^{j} \otimes \rho_{2}^{j}
\end{equation}
where $\rho_{1}^j=|{\bf k}_{j}\rangle_{1\:1} \langle {\bf k}_{j}|$ and $\rho_{2}^j=|-{\bf k}_{j}\rangle_{2\:2}
\langle -{\bf k}_{j}|$ are the density matrices for particles 1 and 2, respectively, belonging to the $j^{th}$
pair. It is well known that for each particle to propagate with such a perfectly well defined momentum, the
sources have to be infinite in the transverse direction \cite{pointsource,Popper_exp}. Therefore, pairs of
particles with a perfect momentum-momentum correlation do not exhibit any position-position correlation. In the
more realistic case of finite transverse dimension of the source, the position-position correlation improves at
the expenses of the momentum-momentum correlation: each particle is always diffracted independently. In general,
any attempt to improve the classical correlation in one variable inevitably worsens the correlation in the
other. Thus the classical statistical inequalities:
\begin{eqnarray}
\Delta (k_1+ k_2) = \sqrt{(\Delta k_1)^{2} + (\Delta k_2)^{2}} > max(\Delta k_1,\Delta k_2), \\ \nonumber \Delta
(x_1- x_2) = \sqrt{(\Delta x_1)^{2} + (\Delta x_2)^{2}} > max(\Delta x_1,\Delta x_2),
\end{eqnarray}
can never be violated by pairs of classically correlated particles/beams.

In conclusion, any source of classically correlated
pairs of particles: i) can never achieve perfect correlation in
both momentum and position variables; ii) can never satisfy
inequalities Eq.~(\ref{ineq}).
%%%%%%%%%%%%%%%%%%%%%%%%%%%%%%%

In summary, we have experimentally demonstrated that SPDC photon pairs satisfy the EPR-type non-classicality
condition Eq.~(\ref{ineq}). In doing so, we have shown that entangled particles exhibit almost perfect $both$
momentum-momentum $and$ position-position correlations. Classically correlated pairs of particles cannot exhibit
such behavior. The measurement described in this Letter thus provides a direct way to distinguish between
quantum entanglement and classical correlation in momentum and/or position variables, for two-photon systems. An
important practical consequence is that only the non-local correlation implicit in entangled systems allows to
`overcome' the usual diffraction limit and to obtain super-resolved images, as proposed and demonstrated in
Ref.~\cite{Dowling,Milena,Popper_exp}. Furthermore, our experiment shows that a distinction between classically
correlated and quantum entangled systems, in momentum and/or position variables, can be realized experimentally
through the study of ``ghost" imaging-type experiments \cite{explain}. This is a quite different approach with
respect to Bell's inequality and may represent an extension of Bell's inequality, in optics.

The authors thank M.H. Rubin and V. Protopopescu for helpful comments and discussions. This work was supported,
in part, by ARDA, NASA-CASPR program, NSF, and ONR. The Oak Ridge National Laboratory is managed by UT-Battelle,
LLC, for the U.S. DOE under contract DE-AC05-00OR22725.


\begin{thebibliography}{}
\bibitem{EPR ineq} Note indeed that in the special case $\Delta(k_1+k_2)=0$,
we get $\Delta(k_1+k_2)\Delta(x_1-x_2)=0$ as well, independently on the value of $\Delta(x_1-x_2)$. However, if
$\Delta(x_1-x_2)>>\Delta(x_2)$, the measurement of $x_1$ would not determine $x_2$ and the EPR argument on the
physical reality of $x_2$ would not apply. We then believe it is formally incorrect to consider
$\Delta(k_1+k_2)\Delta(x_1-x_2)<1$ as the EPR inequality.
\bibitem{explain} Cancellation of all first order effects is a prerequisite for making
such a distinction since quantum `ghost" interference-imaging
phenomena are second order effects.
\end{thebibliography}
\end{document}